\documentclass[journal]{journal}
\usepackage{graphicx}      
\usepackage{amsfonts}
\usepackage{amssymb}
\usepackage{algorithm}
\usepackage{algpseudocode}
\usepackage{mdwmath}
\usepackage{mdwtab}
\usepackage{eqparbox}
\usepackage{url}
\usepackage{physics,amsmath}
\usepackage[T1]{fontenc}
\usepackage{amsmath}
\usepackage{multicol}
\usepackage{flushend}
\usepackage{graphicx}
\usepackage{longtable}
\usepackage{threeparttable}
\usepackage{color}
\usepackage{setspace}
\usepackage{balance}
\usepackage{lastpage}
\def\dis{\displaystyle}

\begin{document}
\title{Quantum State Transfer Optimization: Balancing Fidelity and Energy Consumption using Pontryagin Maximum Principle}

\author{Nahid Binandeh Dehaghani
        and A. Pedro Aguiar
\thanks{The authors gratefully acknowledge the financial support provided by the Foundation for Science and Technology (FCT/MCTES) within the scope of the PhD grant 2021.07608.BD, the Associated Laboratory ARISE (LA/P/0112/2020), the R\&D Unit SYSTEC through Base (UIDB/00147/2020) and Programmatic (UIDP/00147/2020) funds and project RELIABLE - Advances in control design methodologies for safety-critical systems applied to robotics (PTDC/EEI-AUT/3522/2020), all supported by national funds through FCT/MCTES (PIDDAC). The work has been done in honor and memory of Professor Fernando Lobo Pereira.}
}

\markboth{Journal of \LaTeX\ Class Files,~Vol.~6, No.~1, January~2007}%
{Shell \MakeLowercase{\textit{et al.}}: Bare Demo of IEEEtran.cls for Journals}

\maketitle
\thispagestyle{empty}

\begin{abstract}
In this study, we address a control-constrained optimal control problem pertaining to the transformation of quantum states. Our objective is to navigate a quantum system from an initial state to a desired target state while adhering to the principles of the Liouville-von Neumann equation. To achieve this, we introduce a cost functional that balances the dual goals of fidelity maximization and energy consumption minimization.
We derive optimality conditions in the form of the Pontryagin Maximum Principle (PMP) for the matrix-valued dynamics associated with this problem. Subsequently, we present a time-discretized computational scheme designed to solve the optimal control problem. This computational scheme is rooted in an indirect method grounded in the PMP, showcasing its versatility and efficacy.
To illustrate the practicality and applicability of our methodology, we employ it to address the case of a spin $\frac{1}{2}$ particle subjected to interaction with a magnetic field. Our findings shed light on the potential of this approach to tackle complex quantum control scenarios and contribute to the broader field of quantum state transformations.
\end{abstract}


%
\IEEEpeerreviewmaketitle

\section{Introduction}
Quantum technology aspires to develop practical applications based on properties of the systems obeying the laws of quantum mechanics. This objective requires efficient manipulation of quantum objects in order to obtain desired behaviors. Quantum control holds within a number of techniques to obtain the time evolution of control parameters and enable useful performance in applications ranging from quantum computing, \cite{palao2002quantum} to sensing, \cite{rembold2020introduction}, simulation, \cite{holland2020optimal}, and metrology, \cite{lin2021optimal}. Several quantum control methods, such as brute-force optimization of a few pulse parameters, \cite{cheng2020accqoc}, Brumer-Shapiro coherent control, \cite{gruebele2001fully}, pulse-timing control, \cite{mcdermott2014accurate},  stimulated-Raman-Adiabatic-Passage, \cite{vitanov2017stimulated}, genetic algorithms, \cite{lahoz2016quantum}, and optimal control theory (OCT), \cite{werschnik2006quantum}, have been exploited in order to discover the optimal pulse sequences. The studies on Quantum Optimal Control (QOC) began in the late 1980's, \cite{peirce1988optimal}, and has undergone continuous developments up to now.

A wide range of problems arising in quantum technology, as in quantum computing or nuclear magnetic resonance spectroscopy, are compatible to be formulated in the framework of OCT. Amongst the most dominant advances of QOC, we can point out the introduction of rapidly converging iterative algorithms, \cite{maday2003new}, and its generalization for dissipative systems, \cite{ohtsuki1999monotonically}, while taking several control criteria into account. In fact, QOC aims at developing an organized and rigorous design methodology in order to control the behavior of quantum systems so that a desired set of objectives can be obtained in an optimal way. A control field $u(t)$ is able to produce the global maximum or minimum value of a performance index $J(u(t))$, e.g. maximum fidelity, \cite{DEHAGHANI2022214a} or minimum time, \cite{sugny2007time}, while overcoming decoherence and dissipation. 

Despite the considered research works that have been done in quantum optimal control theory, Pontryagin maximum principle of optimal control is still far from being fully exploited in quantum context. In the current state of the art, the quantum systems to be controlled are usually simple closed systems, in which the quantum state is expressed in the form of unit vectors, and evolves according to the Schrodinger equation, \cite{boscain2021introduction}. However, in practical applications, the quantum systems to be controlled are usually not simple closed systems. They may be quantum ensembles, and their states cannot be expressed in the form of unit vectors. In this paper, we address precisely this problem by considering the evolution of the matrix-valued probability density function, and describing the system dynamics in terms of a density operator, by means of, e.g., master equations. As far as the authors' knowledge, such a study has not been done, with exception of our previous recent work in \cite{DEHAGHANI2022214a}. In this paper, we extend and generalize our previous results by applying the Maximum Principle of Pontryagin for the matrix valued quantum dynamic control system, where the state of the quantum system is described through the density operator evolving according to Liouville-von Neumann equation. We aim a trade off between the goal of attaining the maximum fidelity in order to signify a security level for quantum state transformation, which is of high importance notably in quantum information theory, and keep the energy of the field small. The first-order necessary optimality conditions obtained from the application of PMP results in a two-point boundary value problem, which we managed to solve by proposing a shooting algorithm, where we also consider the problem of control constraints. Simulation results illustrate the effectiveness of the proposed methodology. 

The paper is organized as follows: We first review the set of PMP optimality conditions for a general system. Then, we turn to the context of quantum systems, and we study the basis of a quantum optimal control problem, including the current state of the art. Next, we  address the questions of existence of an optimal control and also controllability of a quantum system. Then, we present the physical description of a simple two-level quantum-mechanical system, which we have used in our work. In the next section, we describe the system under study, and, then, we formulate the optimal control problem under control constraints and obtain the necessary conditions of optimality. We later present the application of PMP by means of an indirect method through an algorithm, including simulation results. The paper ends with brief conclusions and an overview on prospective research challenges. 

\subsection{Notation.} 
For a general continuous-time trajectory $x$, the term $x(t)$ indicates the trajectory assessed at a specific time $t$. For writing partial differential equations (PDEs), we denote partial derivatives using subscripts. In the general situation where $f$ denotes a function of $n$ variables including $x$, then $f_x$ denotes the partial derivative relative to the $x$ input. Throughout the paper, we have used $i$ as the imaginary unit. For a matrix $A$, $A^T$
and $A^\dagger$ represent the transpose
and conjugate transpose of matrix A, respectively. To denote the wavefunctions as vectors, we use the Dirac notation such that $\left| \psi  \right\rangle =\sum\limits_{j=1}^{n}{{{\alpha }_{j}}\left| {{{\hat{\psi }}}_{j}} \right\rangle }$, where $\vert \psi \rangle $ indicates a state vector, ${\alpha }_{j}$ are the complex-valued expansion coefficients, and $\vert {\hat \psi}_j \rangle $ are basis vectors that are fixed.
\newline
We denote a finite-dimensional separable Hilbert space by $\mathbb{H}$, and define it over the complex field $\mathbb{C}$, so $\mathbb{H}\simeq\mathbb{C}^N$, where $N$ indicates the dimension of the space. We consider the set $\mathcal{B}(\mathbb{H})$ as the set of linear operators on the Hilbert space and define it as $\mathcal{B}(\mathbb{H})\simeq\mathbb{C}^{N \times N}$. 
The inner product in the set $\mathcal{B}(\mathbb{H})$ is the Hilbert-Schmidt inner product, defined as $\left\langle A,B \right\rangle =tr\left( {{A}^{\dagger }}B \right)$, where $tr(A)$ indicates the trace of an square matrix $A$. The commutator of two elements $A$ and $B$, as linear operators on the Hilbert space, is indicated by $[A,B]:=AB-BA$.

\section{Pontryagin's maximum principle}
In general terms, a fundamental optimal control problem is formulated as the following \cite{werschnik2006quantum}:

Given the dynamical system $\dot{x}\left( t \right)=f\left( x\left( t \right),u\left( t \right) \right)$
and a set of admissible controls $u(t)\in \mathcal{U}$, we have to determine an admissible control signal such that the objective functional 
\begin{equation}\label{1}
J=\Phi(x(T))+\int\limits_{0}^{T}{L}\left( x\left( t \right),u\left( t \right) \right){dt}  \end{equation}
is minimized. Let $u^*(t)\in \mathcal{U}$ and $x^*(t)$ represent the optimal control and state trajectory for the defined optimal control problem. Then, there is a time-varying adjoint trajectory $\lambda(t)$ that together with $u^*(t)$ and $x^*(t)$ satisfy 

\begin{equation}\label{2}
\begin{aligned}
&\text{System Equation and Initial State Condition} && \\
&\dot{x}^*(t) = f(x^*(t), u^*(t)) && \\
&x^*(t=0) = x_0 &&
\end{aligned}
\end{equation}

\begin{equation}\label{3}
\begin{aligned}
&\text{Adjoint Equation and Transversality Condition} && \\
&-{\dot{\lambda}}^\dagger(t) = {\lambda}^\dagger(t) f_x(x^*(t), u^*(t)) - L_x(x^*(t), u^*(t)) && \\
&{\lambda}^\dagger(T) = -\Phi_x(x^*(T)) &&
\end{aligned}
\end{equation}

\begin{equation}\label{4}
\begin{aligned}
&\text{Maximum Condition} && \\
&\mathcal{H}(\lambda(t), x^*(t), u(t)) \le \mathcal{H}(\lambda(t), x^*(t), u^*(t)) && \\
&&
\end{aligned}
\end{equation}
where $\mathcal{H}$ is the Pontryagin Hamiltonian defined as
\small
\begin{equation}\label{5}
\! \! \!   \mathcal{H}\left( \lambda \left( t \right),x\left( t \right),u\left( t \right) \right):={{\lambda }^{\dagger }}\left( t \right)f\left( x\left( t \right),u\left( t \right) \right)-L\left( x\left( t \right),u\left( t \right) \right)
\end{equation}
\normalsize
\section{An overview on quantum optimal control}
Quantum Optimal Control can intuitively be formulated in the above-mentioned setting. For such problems, the state of the system may be described by the pure state vector, density operator (for both mixed and pure quantum states), or we can consider the dynamics of the evolution operator.
The evolution of a pure state, which is not entangled with the environment, can be described by a wave function $\left| \psi (t) \right\rangle$, evolving in time according to a control-dependent Schrodinger equation,  \cite{boscain2021introduction}, 
\begin{equation}\label{6}
	i\hbar{ \left| \dot{\psi }\left( t \right) \right\rangle }=H\left( u\left( t \right) \right)\left| \psi \left( t \right) \right\rangle, \quad \left| \psi (t=0) \right\rangle =\left| {{\psi }_{0}} \right\rangle
\end{equation}
where ${H(u(t))}$ is the quantum-mechanical Hamiltonian of the system, and $\hbar$ is the reduced Planck constant, usually set as $\hbar=1$ for convenience. The system control can be realized by an admissible set of external control signals ${{u}_{k}}(t)\in \mathbb{R}$, which are coupled to the quantum system via time independent interaction Hamiltonians ${{H}_{k}}$. Therefore, the total quantum Hamiltonian defined as 
\begin{equation}\label{7}
 H\left(u\left( t \right)\right)={{H}_{0}}+\sum\limits_{k=1}^m{{{u}_{k}}(t)}{{H}_{k}}   
\end{equation}
determines the controlled evolution, in which ${{H}_{0}}$ indicates the time independent internal (free) Hamiltonian.
In \eqref{6}, both the system state and Hamiltonian are complex quantities. However, we can express the problem in terms of only real quantities by introducing $x:={{\left[ \vec{\psi}_{R}^{T},\vec{\psi }_{I}^{T} \right]}^{T}}$, where $\left| \psi_R (t) \right\rangle, \left| \psi_I (t) \right\rangle \in \mathbb{R}^n$, and separating the real and imaginary parts of $-iH\left( u(t) \right)=R\left( u(t) \right)+iI\left( u(t) \right)$, where $R\left( u(t) \right), I\left( u(t) \right)\in {{\mathbb{R}}^{n\times n}}$ are skew-symmetric and symmetric matrices for all values of $u(t)$, respectively. Now, we can rewrite the differential equation describing the dynamics of the system implying only real values by 
\begin{equation}\label{8}
\dot{x}=\tilde{H}\left( u(t)  \right)x
\end{equation}
in which
\begin{equation}\label{9}
    \tilde{H}\left( u(t)  \right)=\left( \begin{matrix}
   R\left( u(t)  \right) & -I\left( u(t)  \right)  \\
   I\left( u(t)  \right) & R\left( u(t)  \right)  \\
\end{matrix} \right)
\end{equation}
is both symplectic and skew-symmetric for all values of $u(t)$. The cost in \eqref{1} can also be rewritten by introducing appropriate functions $\tilde{\Phi}$ and $\tilde{L}$. Hence, the Pontryagin Hamiltonian takes the form 
\begin{equation}\label{10}
\!\!\!\!\!\! \mathcal{H}\left( \lambda \left( t \right),x\left( t \right),u\left( t \right) \right)={{\lambda }^{T}}\left( t \right)\tilde{H}\left( u\left( t \right) \right)x\!-\!\tilde{L}\left( x\left( t \right),u\left( t \right) \right) \!
\end{equation}
from which the optimal control $u(t)$ has to satisfy the maximum condition \eqref{4}. Since $\tilde{H}\left( u(t) \right)$ is skew-symmetric, \eqref{3} can be rewritten as \cite{dehaghani2022optimal},
\begin{equation}\label{11}
\begin{aligned}
  & \dot{\lambda }\left( t \right)=\tilde{H}\left( u \right)\lambda \left( t \right)+\tilde{L}_{x}^{T} \\ 
 & \lambda \left( T \right)=-\tilde{\Phi }_{x}^{T}\left( x(T) \right) \\ 
\end{aligned}
\end{equation}

\section{Existence of the Optimal Control}
Since we use the necessary optimality conditions, we also need to address the problem of existence of the optimal control. There are already some standard results, well explained in \cite[Chapter III]{fleming2012deterministic}, that are applicable to the case considered in our work. According to \cite{fleming2012deterministic}, the existence theorem guarantee the existence of an optimal control in the Lebesgue Integrable ($LI$) set if a control steering a given initial state to a desired target state exists. The control functions $u_k(t)$ in \eqref{8} are supposed to be $LI$ as well. We also will suppose in this paper that the optimal control are in the set $LI$, according to the above-mentioned theorem. In order to guarantee the existence of a control, one needs to address the question of controllability. The controllability concerns the possibility of steering the system from one state to another for every pair of states. 

The complex sphere $\mathbb{S}^{2N-1}\subset \mathbb{H}$, representing the pure quantum states, is a homogeneous space of the Lie group
$U\left( N \right)=\left\{ U\in GL\left( N,\mathbb{C} \right)\left| U{{U}^{\dagger }} \right.={{U}^{\dagger }}U=I \right\}$, and its proper subgroup $SU\left( N \right)=U\left( N \right)/U\left( 1 \right)$. The Lie algebras of $U(N)$ and $SU(N)$ are 
\begin{equation}\label{12}
u\left( N \right)=\left\{ A\in {{\mathbb{C}}^{N\times N}}\left| {{A}^{\dagger }}=-A \right. \right\} 
\end{equation}
and
\begin{equation}\label{13}
 su\left( N \right)=\left\{ A\in u\left( N \right)\left| tr\left( A \right)=0 \right. \right\},   
\end{equation}
respectively. The Schrodinger equation stated in \eqref{6} can be lifted to the Lie group $SU(N)$ to obtain the Schrodinger equation for the unitary propagator as
\begin{equation}\label{14}
    i\frac{d}{dt}U\left( t \right)=\left( {{H}_{0}}+\sum\limits_{k=1}^{m}{{{u}_{k}}\left( t \right){{H}_{k}}} \right)U\left( t \right), \quad U(0)=I
\end{equation}
where $U\in SU(N)$. Equation \eqref{14} is a right invariant control system on the compact Lie group $su(n)$, so
\begin{equation}\label{15}
Lie\left\{ i{{H}_{0}},\ldots ,i{{H}_{m}} \right\}=su\left( n \right)
\end{equation}
which represents a necessary and sufficient condition for the controllability of the system indicated in \eqref{14}, \cite{boothby1979determination}. Overall, if a right invariant system is controllable, then the bilinear system is also controllable, \cite{sachkov1997controllability}. Therefore, if the evolution operator of the system satisfies a controllable equation, then the bilinear system described by the certain quantum-mechanical Hamiltonian $H\left(u\left( t \right)\right)$ is controllable as well, so it is possible to design controls to steer the bilinear system from one state to another state in the state-space. Consequently, the system indicated in \eqref{6} is controllable provided that \eqref{15} holds. The controllability of two-level quantum systems, as the case considered in this paper, is analysed in more details in \cite{d2000topological}.

\section{Physical Description of a Control System}
In Nuclear Magnetic Resonance (NMR) experiments, a single spin $\frac{1}{2}$ particle is controlled by means of an electromagnetic field $\vec{B}\left( t \right)$, in which one component is kept constant in $z$ direction while $x$ and $y$ components vary in time in order to change the direction of the spin, \cite{d2021introduction}. The total quantum mechanical Hamiltonian is then described by the interaction of the external magnetic field $\vec{B}\left( t \right)=\left({{B}_{x}}\left( t \right),{{B}_{y}}\left( t \right),{B}_{z} \right)^T$ with the spin angular momentum $\hat{S}:=\left({{{\hat{S}}}_{x}},{{{\hat{S}}}_{y}},{{{\hat{S}}}_{z}} \right)^T$ as
\begin{equation} \label{16}
\begin{aligned}
 H\left( t \right)&=\gamma \hat{S}^T\vec{B}\left( t \right) \\ 
 & =\gamma \left( {{{\hat{S}}}_{x}}{{B}_{x}}\left( t \right)+{{{\hat{S}}}_{y}}{{B}_{y}}\left( t \right)+{{{\hat{S}}}_{z}}{{B}_{z}} \right) \\ 
\end{aligned}
\end{equation}
where $\gamma$ is the gyromagnetic ratio. 
Let express the state $\left| \psi \left( t \right) \right\rangle ={{c}_{1}}\left( t \right)\left| \frac{1}{2} \right\rangle +{{c}_{2}}\left( t \right)\left| -\frac{1}{2} \right\rangle$. From \eqref{6} and \eqref{16}, the differential equation for $\vec{c}\left( t \right)={{\left( {{c}_{1}}\left( t \right),{{c}_{2}}\left( t \right) \right)}^{T}}$ is expressed as
\begin{equation}\label{17}
    i\frac{d}{dt}\vec{c}= \frac{\gamma}{2} \left( {{\sigma }_{x}}{{B}_{x}}\left( t \right)+{{\sigma }_{y}}{{B}_{y}}\left( t \right)+{{\sigma }_{z}}{{B}_{z}} \right)\vec{c}
\end{equation}
in which the matrix representation of the operators $\sigma_{x}$, $\sigma_{y}$, and $\sigma_{z}$ is done by the so-called Pauli matrices. By appropriate scaling of time and setting the magnetic field arguments as controls we have
\begin{equation}\label{18}
    -iH\left( u\left( t \right) \right)=\left( {{{\bar{\sigma }}}_{z}}{{u}_{z}}+{{{\bar{\sigma }}}_{x}}{{u}_{x}}\left( t \right)+{{{\bar{\sigma }}}_{y}}{{u}_{y}}\left( t \right) \right)
\end{equation}
where 
\begin{equation*}
\! \!
    {{\bar{\sigma }}_{x}}=\frac{1}{2}\left( \begin{matrix}
   0 & i  \\
   i & 0  \\
\end{matrix} \right),\! \! \quad {{\bar{\sigma }}_{y}}=\frac{1}{2} \left( \begin{matrix}
   0 & -1  \\
   1 & 0  \\
\end{matrix} \right),\! \! \quad {{\bar{\sigma }}_{z}}=\frac{1}{2} \left( \begin{matrix}
   i & 0  \\
   0 & -i  \\
\end{matrix} \right) \!
\end{equation*}
span $SU(2)$ of skew-Hermitian matrices with trace equal to zero, and satisfy the following commutation relations
\begin{equation}\label{19}
    \left[ {{{\bar{\sigma }}}_{x}},{{{\bar{\sigma }}}_{y}} \right]={{\bar{\sigma }}_{z}},\quad\left[ {{{\bar{\sigma }}}_{y}},{{{\bar{\sigma }}}_{z}} \right]={{\bar{\sigma }}_{x}},\quad\left[ {{{\bar{\sigma }}}_{z}},{{{\bar{\sigma }}}_{x}} \right]={{\bar{\sigma }}_{y}}\cdot
\end{equation}
Equation \eqref{16} can be implemented according to \eqref{8}, to obtain 
\begin{equation}\label{20}
    	\tilde{H}\left( u\left( t \right) \right)={{T}_{z}}{{u}_{z}}+{{T}_{y}}{{u}_{y}}\left( t \right)+{{T}_{x}}{{u}_{x}}\left( t \right)
\end{equation}
where 
{\small
\begin{equation}\label{21}
\begin{aligned}
  & {{T}_{x}}=\frac{1}{2}\left( \begin{matrix}
   0 & 0 & 0 & -1  \\
   0 & 0 & -1 & 0  \\
   0 & 1 & 0 & 0  \\
   1 & 0 & 0 & 0  \\
\end{matrix} \right),{{T}_{y}}=\frac{1}{2}\left( \begin{matrix}
   0 & -1 & 0 & 0  \\
   1 & 0 & 0 & 0  \\
   0 & 0 & 0 & -1  \\
   0 & 0 & 1 & 0  \\
\end{matrix} \right), \\ 
 & {{T}_{z}}=\frac{1}{2}\left( \begin{matrix}
   0 & 0 & -1 & 0  \\
   0 & 0 & 0 & 1  \\
   1 & 0 & 0 & 0  \\
   0 & -1 & 0 & 0  \\
\end{matrix} \right) \cdot \\ 
\end{aligned}
\end{equation} } 
From now on, we show the two components of the time varying control by the vector $\vectorbold{u}(t)=\left[ \begin{matrix}
   {{u}_{x}}\left( t \right) & {{u}_{y}}\left( t \right)  \\
\end{matrix} \right]^T$. 

\section{System Description}
Every state vector $\left| {{\psi }_{j}}\left( t \right) \right\rangle $ at time $t$ can be obtained by
\begin{equation}\label{22}
    \left| {{\psi }_{j}}\left( t \right) \right\rangle=U(t)\left| {{\psi }_{j}}\left( 0 \right) \right\rangle
\end{equation} 
in which $U(t)$ is the solution of \eqref{14} expressed as
\begin{equation}\label{23}
   U\left( t \right)=\exp \left( -i\int\limits_{0}^{t}{H\left( s \right)ds} \right) 
\end{equation}
called Dyson's series in the physics context and is similar to a Volterra series in control theory, \cite{altafini2012modeling}. Hence, the outer product $\left| {{\psi }_{j}}\left( t \right) \right\rangle \left\langle  {{\psi }_{j}}\left( t \right) \right|$ results in 
\begin{equation}\label{24}
\left| {{\psi }_{j}}\left( t \right) \right\rangle \left\langle  {{\psi }_{j}}\left( t \right) \right|=U\left( t \right)\left| {{\psi }_{j}}\left( 0 \right) \right\rangle \left\langle  {{\psi }_{j}}\left( 0 \right) \right|{{U}^{\dagger }}\left( t \right)
\end{equation}
and the same for any convex sum of $\left| {{\psi }_{j}}\left( t \right) \right\rangle \left\langle  {{\psi }_{j}}\left( t \right) \right|$. More precisely, let $p_j$ be the fraction of population of an ensemble 
$\left\{ \left. {{p}_{j}},\left| {{\psi }_{j}}\left( t \right) \right\rangle  \right\} \right.$, so the corresponding quantum density operator is expressed as 
\begin{equation}\label{25}
\rho =\sum\limits_{j}{{{p}_{j}}}\left| {{\psi }_{j}} \right\rangle \left\langle  {{\psi }_{j}} \right|,\quad{{p}_{j}}\ge 0,\quad \sum\limits_{j}{{{p}_{j}}}=1    
\end{equation}
which belongs to the set of Hermitian, semi definite, and positive matrices with trace equal to one on the system?s Hilbert space $\mathbb{H}$. Hence, we can rewrite \eqref{25} as
\begin{equation}\label{26}
    \rho \left( t \right)=U\left( t \right)\rho \left( 0 \right){{U}^{\dagger }}\left( t \right)
\end{equation}
The infinitesimal version of \eqref{26} is the quantum Liouville-von Neumann equation, \cite{berman1991time}, expressed by
\begin{equation}\label{27}
\dot{\rho }\left( t \right)=-i\left[ H\left( u\left( t \right) \right),\rho \left( t \right) \right], \quad \rho(0)=\rho_0 
\end{equation}
The controllability of \eqref{27} can consequently be obtained from the necessary and sufficient condition indicated in \eqref{15}.
The main characteristic of the Liouville-von Neumann equation is that it generates isospectral evolutions, meaning that
\begin{equation}
    sp\left( \rho \left( t \right) \right)=sp\left( \rho \left( 0 \right) \right)=\Phi \left( \rho  \right)=\left\{ {{\mu }_{1}},\ldots ,{{\mu }_{N}} \right\}
\end{equation}
where $\mu_1,\dots,\mu_N$ are the eigenvalues of $\rho(t)$. As a consequence of the isospectrality of \eqref{27}, the set $\Phi \left( \rho  \right)$ forms a complete set of constants of motion of \eqref{27}. Let consider the set $\mathcal{D}\left( \mathbb{H} \right)=\left\{ \rho \in \mathcal{B}\left( \mathbb{H})\right)\left| \rho ={{\rho }^{\dagger }}\ge 0,tr\left( \rho  \right)=1 \right. \right\}$, which is foliated into leaves uniquely determined through the set $\Phi(\rho)$, and $\mathcal{C} \subset \mathcal{D}\left( \mathbb{H} \right)$ as such as leaf, where $\rho_0 \in \mathcal{C} $. Hence, $\mathcal{C}=\left\{ U{{\rho }_{0}}{{U}^{\dagger }},U\in SU\left( N \right) \right\}$ corresponds to the orbit of $SU(N)$ under the action of conjugation passing through the initial density operator. Assume $j_1,\dots,j_l$ as the geometric multiplicities of the eigenvalues $\Phi(\rho_0)$, and $j_1+\dots+j_l=N$, $2\le l \le N$, then $\mathcal{C}$ is the homogeneous space as $C=U\left( N \right)/\left( U\left( {{j}_{1}} \right)\times \ldots \times U\left( {{j}_{l}} \right) \right)$. As the eigenvalues of the density operator vary, the geometric multiplicities $j_1,\dots,j_l$ form a flag, so the $\mathcal{C}$ are called complex flag manifolds, \cite{bengtsson2017geometry}. The flag also determines the dimension of flag manifolds $\mathcal{C}$, varying from $2N-2$, for pure states, to $N^2-N$, for all various eigenvalues, \cite{altafini2012modeling}. Equation \eqref{27} will be adopted as the investigated model in the following subsections. 
\section{Formulation of the Optimal Control Problem}
In optimal control of quantum state transfer problems, a possible cost in the form of \eqref{1} to be minimized is to consider a trade off between the goal of attaining the maximum fidelity in order to signify a security level for quantum state transformation, while simultaneously keeping the energy of the field small. Motivated by this consideration, we propose the following optimal control problem $(P_1)$
\begin{eqnarray*}
   \mbox{Minimize} \\
    && J=-\mathcal{F}(\rho(T),\sigma )+\eta \int\limits_{0}^{T}{
    \vectorbold{u}^{T}(t)\vectorbold{u}(t)}dt \\
    &&  t\in [0,T],\quad \eta \in [0,1]
    \end{eqnarray*}
    \begin{eqnarray*}
    \mbox{Subject to }
    \\
    &&\textit{dynamics}: \dot{\rho}(t)=F\left( \rho(t),u(t) \right),\\
    && \qquad \qquad \quad  \rho\left( t \right)\in {{\mathbb{C}}^{n\times n}} \\
    &&\textit{initial condition}: \rho\left( 0 \right)={{\rho}_{0}} \\
    &&\textit{control constraint}: u(\cdot) \in \mathcal{U}, \textit{i.e., }\\
    && \mathbf{u}\in L^2(0,T)^2:\textit{ for a.e. } t\in[0,T]\\
    &&-u_{max}\le u_x(t)\le u_{max},\\
    &&-u_{max}\le u_y(t)\le u_{max}, \\
\end{eqnarray*}
The coefficient $\eta\ge 0$ is considered to signify the importance of energy minimization. The density operator $\rho(t)$ is the quantum state variable supposed to satisfy the differential constraints according to the Liouville-von Neumann equation \eqref{27}, and $\rho_0$ is the so-called initial quantum state. The set of admissible controls $\Omega$ is defined as
\begin{equation*}
 \Omega =\left\{ \left[ \begin{matrix}
   {{u}_{x}}(t) & {{u}_{y}}(t)   \\
\end{matrix} \right]{{\left| \left\| u \right\| \right.}^{2}}\le \sqrt{2}{{u}_{\max }} \right\}   
\end{equation*}
where ${{u}_{\max }}$ is a given positive parameter. In P1, we have used the well-known Uhlmann-Jozsa definition for fidelity representing the maximal transition probability between the purification of a pair of density matrices, $\rho(T)$ and the desired target state $\sigma$, \cite{liang2019quantum}, defined as
    \begin{equation}\label{28}
        \mathcal{F}\left( \rho(T) ,\sigma  \right):={{\left( tr\sqrt{\sqrt{\rho(T) }\sigma \sqrt{\rho(T) }} \right)}^{2}}\cdot
    \end{equation}
    
\section{Necessary Conditions of Optimality in the form of a Maximum Principle}

The Pontryagin-Hamilton function ${\cal H}$ is defined for almost all $t\in [0,T]$ by introducing the matrix-valued time-varying multiplier $\Lambda$, designated by costate or adjoint variable of the system, \cite{DEHAGHANI2022214a}.
Thus,
\begin{equation}\label{29}
    {\cal H}\left( \rho ,u,\Lambda \right)=tr\left( \Lambda^\dagger F\left( \rho ,u \right) \right)-L(u(t))
\end{equation}
According to the Pontryagin's Maximum Principle \eqref{4}, for the optimal state trajectory $\rho^*$ and the corresponding adjoint variable $\Lambda$, the optimal control $u^*(t)$ maximizes the Pontryagin-Hamiltonian function ${\cal H}$ for almost all $t\in[0,T]$ and all admissible control values $u \in \Omega$ such that
\begin{equation}\label{30}
 {\cal H}\left( \rho^*(t), u(t), \Lambda(t) \right)\leq {\cal H}\left( \rho^*(t), u^*(t),\Lambda (t) \right)
\end{equation}
Consequently, the adjoint equation implies that
\begin{equation}\label{31}
\begin{aligned}
-\dot\Lambda^\dagger(t)&={\cal H}_{\rho}( \rho^*(t) ,u^*(t), \Lambda(t))\\
&=i\left[ H( u^*(t) ),\Lambda^\dagger(t) \right]\\
\end{aligned}
\end{equation}
in which $H^*(t)$ denotes the quantum mechanical Hamiltonian evaluated at each time along the optimal control $u^*$. Equation \eqref{31} has the formal solution
\begin{equation}\label{32}
    \Lambda^\dagger( t)=e^{i\int_t^TH^*(s)ds}\Lambda^\dagger(T)e^{-i\int_t^TH^*(s)ds}\cdot
\end{equation}
The boundary condition at the final time for adjoint variable implies that
\begin{equation}\label{33}
\begin{aligned}
\Lambda^\dagger( T)&=\nabla_\rho\left( tr\sqrt{\sqrt{\rho(T) }\sigma \sqrt{\rho (T) }}\right)^2\\
&=2 tr\sqrt{\rho(T) \sigma}\sum\limits_{k=0}^{n-1}{\alpha }_{k}\sum_{i=0}^{k-1}\bar \rho(T)^i\sqrt{\sigma}\bar \rho(T)^{k-i-1}
\end{aligned}
\end{equation}
in which $n$ indicates the dimension of the density matrix and $\bar\rho(T)= \rho(T)-I$. The derivation of \eqref{33} can be found in \cite{DEHAGHANI2022214a}, where the coefficients $\alpha_k$, $k=0,\ldots, n-1$ are obtained from the application of the Cayley-Hamilton theorem.

\section{Application of the Pontryagin Maximum Principle}

In this section, we solve the optimal control problem defined in $(P_1)$ by means of an indirect method based on the PMP. In the presented algorithm, we have discretized the time interval $[0, T]$ into $N$ sub intervals such that $\dis t_k = \frac{k}{N}$ for $ k = 0,\ldots, N-1$, and setting $j=0,\ldots$ as the iterations counter. Hence, $j^{th}$ iteration of the function $f$ at time $t_k$ is represented by $f_k^j$. Here, both the dynamics of the system and adjoint equation are considered by a first order Euler approximation. The proposed algorithm is explained in \ref{alg}.

\begin{algorithm}\label{alg}
\small
\caption{Optimization Algorithm}
\begin{algorithmic}[1]
\State \textbf{Step 1 - Initialization}:
  \State Initialize the values of $\text{u}_{x_k^j}$ and $\text{u}_{y_k^j}$ for $k=0,\ldots,N-1$, and set $j=1$.

\State \textbf{Step 2 - Computation of the State Trajectory}:
  \For{$k=0$ to $N-1$}
    \State Compute $U_{k}^{j} = e^{\sum\limits_{l=0}^{k-1}{\frac{T}{N}\tilde{H}_{l}^{j}}}$
    \State Obtain $\rho _{k}^{j} = U_{k}^{j}\rho _{0}^{j}U_{k}^{j\dagger}$
  \EndFor

\State \textbf{Step 3 - Computation of the Adjoint Trajectory}:
  \State Compute $\Lambda_N^j$ using $\rho _N^j$ computed in Step 2.
  \For{$k=0$ to $N-1$}
    \State Compute $V_{k}^{j} = e^{\sum\limits_{k}^{N-1}{\frac{T}{N}\tilde{H}_{k}^{j}}}$
    \State Obtain $\Lambda_{k}^{j \dagger} = V_{k}^{j\dagger}\Lambda_{N}^{j\dagger}V_{k}^{j }$
  \EndFor

\State \textbf{Step 4 - Computation of the Pontryagin Hamilton Function}:
  \For{$k=0$ to $N-1$}
    \State Let $\mathcal{H}_k^j(u_k^j) = \mathrm{tr}\left({\Lambda^j_k}^\dagger F_k^j (\rho_k^j,u_k^j)\right) - L^j_k$
  \EndFor

\State \textbf{Step 5 - Computation of the Control Function}:
  \For{$k=0$ to $N-1$}
    \State Compute temporary control values $\text{u}_{\text{temp},x_k^j}$ and $\text{u}_{\text{temp},y_k^j}$ that maximize the map
    \State $({u_x},{u_y}) \to \mathcal{H}_k^j({u_x},{u_y})$
  \EndFor
  
\State \textbf{Step 6 - Apply the Control Constraints}:
  \For{$k=0$ to $N-1$}
    \State $\text{u}_{x_k^j} = \min(|\text{u}_{\text{temp},x_k^j}|, u_{\text{max}}) \cdot \mathrm{sign}(\text{u}_{\text{temp},x_k^j})$
    \State $\text{u}_{y_k^j} = \min(|\text{u}_{\text{temp},y_k^j}|, u_{\text{max}}) \cdot \mathrm{sign}(\text{u}_{\text{temp},y_k^j})$
  \EndFor

\State \textbf{Step 7 - Stopping Test}:
  \State For a determined tolerance error $\varepsilon > 0$, check if:
  \State $\max_{k=0,\ldots,N-1}\{|\text{u}_{x_k^j}-\text{u}_{x_k^{j-1}} |\} < \varepsilon$
  \State $\max_{k=0,\ldots,N-1}\{|\text{u}_{y_k^j}-\text{u}_{y_k^{j-1}} |\} < \varepsilon$

  \If{convergence criteria met}
    \State Let $\text{u}_{{x}^*(t_k)} = \text{u}_{x_k^j}$ and $\text{u}_{{y}^*(t_k)} = \text{u}_{y_k^j}$ for $k=0,\ldots,N-1$, and exit the algorithm.
  \Else
    \State Update temporary control values:
    \State $\text{u}_{\text{temp},x_k^j} = \text{u}_{x_k^{j-1}} + \delta(\text{u}_{x_k^j} - \text{u}_{x_k^{j-1}})$
    \State $\text{u}_{\text{temp},y_k^j} = \text{u}_{y_k^{j-1}} + \delta(\text{u}_{y_k^j} - \text{u}_{y_k^{j-1}})$
    \State Check the control constraints according to \textbf{Step 6}.
    \State Increment $j$: $j = j + 1$, and go to \textbf{Step 2}.
  \EndIf
\end{algorithmic}
\end{algorithm}
Here, we check the algorithm convergence by verifying whether the control functions obtained in the current iteration approach the ones of the previous iteration with an acceptable tolerance or not. If not, we repeat the above steps until we achieve the desired convergence.

\section{Simulation Results}
\begin{figure}
	\centering
	\includegraphics[scale=0.4]{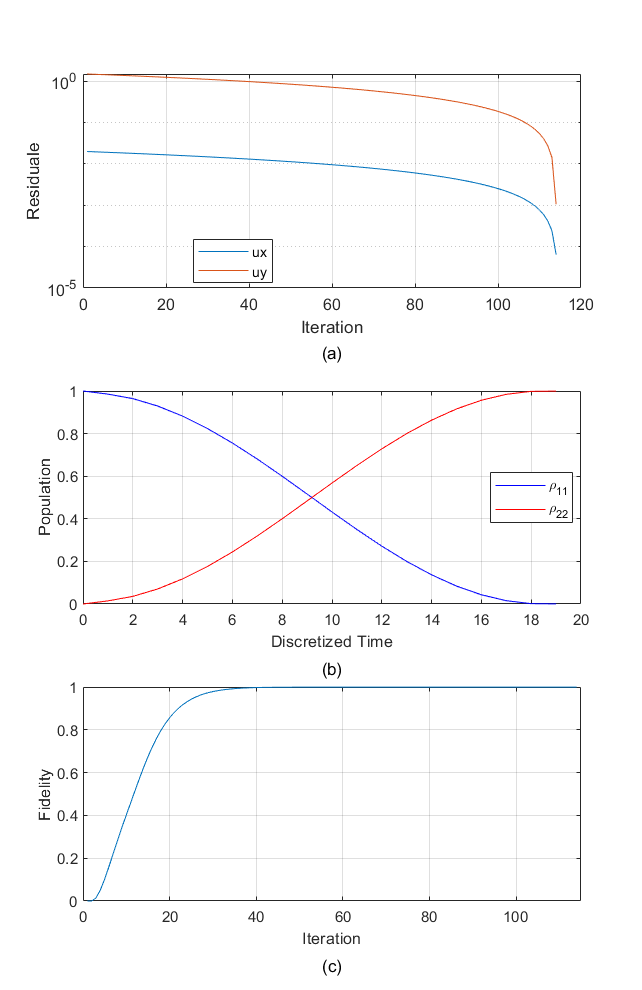}
        \caption{(a) Convergence plot of residuals shows the algorithm converges after 114 iterations. (b) Evolution of diagonal elements of the State trajectory at final iteration over the discretized time. (c) State transition probability.} 
        \label{fig}
\end{figure}

Let consider the quantum state transfer problem given the initial state $\rho_0=\left( \begin{matrix}1 & 0  \\ 0 & 0  \\ \end{matrix} \right)$ and the desired target state $\sigma=\left( \begin{matrix} 0 & 0  \\ 0 & 1  \\ \end{matrix} \right)$. By means of implementing the algorithm explained in the previous section, we solve the quantum state transfer problem $P_1$ to drive the given initial quantum state to a desired set while simultaneously minimizing the required control power with the factor $\eta=0.01$. In this simulation, the control is constrained to $u_x, u_y\in[-1,1]$, and the algorithm has been considered to evolve in $t\in[0,1]$ to accomplish the desired fidelity. The control signal in $z$ direction is considered constant as $u_z=0.001$, and in $x$ and $y$ directions is initialized by ${u_x}_k^j=0$ and ${u_y}_k^j=0$ for $k=0,\ldots,N-1$. By setting the time-slicing $N=20$, the learning coefficient $\delta=0.1$, and the stopping threshold $\varepsilon=10^{-3}$, the algorithm converges after 114 iterations. The residual graph showing the convergence of the algorithm is displayed in Fig.~\ref{fig}(a). 
In Fig.~\ref{fig}(b), 
the evolution of the 
diagonal elements of the density matrix during the time in the last iteration is presented. As seen, a smooth evolution targeting the desired state is obtained at the final iteration. The density matrix has unit trace and is Hermitian during its evolution over the time. The resulted final state is $\rho \left( T \right)=\left( \begin{matrix}
   0.0005 & \text{0}\text{.0157 + 0}\text{.0145i}  \\
   \text{0}\text{.0157- 0}\text{.0145i} & \text{0}\text{.9995}  \\
\end{matrix} \right)$. 


%
The evolution of fidelity along the iterations is demonstrated in Fig.~\ref{fig}(c). The graph shows the fidelity reaches 99.8 percent after 44 iterations, and it continues to reach at more than 99.9 percent, which proves that the algorithm has performed well in optimizing a cost functional combining fidelity and control power required to drive the state of the system from a given point to the given final target state. 
\section{Conclusion}
In this paper, we have demonstrated the practical application of the Maximum Principle of Pontryagin, showcasing its utility in calculating constrained optimal control solutions. Our primary focus has been striking a delicate balance between the twin objectives of maximizing fidelity and maintaining the energy of the quantum field at a manageable level.
Within the context of the explored optimal control problem, we have characterized the quantum system's state using the density operator, which evolves in accordance with the Liouville-von Neumann equation. As a pioneering approach in quantum optimal control, we have derived the essential first-order optimality conditions for matrix-valued dynamics.
Our application of the Maximum Principle of Pontryagin has culminated in the development of the proposed shooting algorithm, a valuable tool employed to tackle the intricacies of the two-point boundary value problem. Notably, this methodology seamlessly extends its applicability to scenarios involving pure state vectors or even the dynamics of the evolution operator.
Looking ahead, our work paves the way for exciting future prospects. These include the exploration of the optimal control paradigm's adaptability to address challenges posed by open quantum systems and the incorporation of additional constraints, such as state constraints. These endeavors promise to further enrich our understanding and harness the full potential of quantum optimal control methodologies.

\end{document}